\documentclass[conference, a4paper]{IEEEtran}
\usepackage[cmex10]{amsmath}
\usepackage{amssymb}
\usepackage{booktabs}
\usepackage{latexsym}
\usepackage{amsfonts}
\usepackage{psfrag}
\usepackage{graphicx}
\usepackage{subfigure}
\usepackage{algorithm}
\usepackage{algorithmic}
\usepackage{balance}
\usepackage{cite}
\usepackage[left=1.31cm, right=1.31cm, top=1.92cm, bottom=3.5cm]{geometry}
\usepackage{url}

\begin{document}

\title{{\it Castor}: Contextual IoT Time Series Data and Model Management at Scale}
\author{ Bei Chen,~Bradley Eck,~Francesco Fusco,~Robert Gormally,~Mark Purcell,~Mathieu Sinn,~Seshu Tirupathi \\ IBM Research -- Ireland \\
 beichen2,~bradley.eck,~francfus,~robertgo,~markpurcell,~mathsinn,~seshutir
@ie.ibm.com
}

\IEEEoverridecommandlockouts
\IEEEpubid{\makebox[\columnwidth]{\copyright2018 IEEE \hfill} \hspace{\columnsep}\makebox[\columnwidth]{}}
\maketitle
\IEEEpubidadjcol

\begin{abstract}
We demonstrate {\it Castor}, a cloud-based system for contextual IoT time series data and model management at scale. {\it Castor} is designed to assist Data Scientists in (a) exploring and retrieving all relevant time series and contextual information that is required for their predictive modelling tasks; (b) seamlessly storing and deploying their predictive models in a cloud production environment; (c) monitoring the performance of all predictive models in production and (semi-)automatically retraining them in case of performance deterioration. 
The main features of {\it Castor} are: (1) an efficient pipeline for ingesting IoT time series data in real time; (2) a scalable, hybrid data management service for both time series and contextual data; (3) a versatile semantic model for contextual information which can be easily adapted to different application domains; (4) an abstract framework for developing and storing predictive models in R or Python; (5) deployment services which automatically train and/or score predictive models upon user-defined conditions. 
We demonstrate {\it Castor} for a real-world Smart Grid use case and discuss how it can be adapted to other application domains such as Smart Buildings, Telecommunications, Retail or Manufacturing.
\end{abstract}

\IEEEpeerreviewmaketitle

\section{Introduction}
One of the major challenges in Internet of Things (IoT) applications is the effective management of large amounts of time series data and associated predictive models. Besides the sheer volume of data, substantial complexity arises from the heterogeneity of data sources: without detailed semantic context information about the type of signals, entities and their geographical location, it is impossible for users (Data Scientists in particular) to make sense of the raw time series data. Even more complexity is introduced by considering predictive models which aim to capture regularities and statistical patterns of time series, often as a function of other time series that are correlated because of co-location and/or physical processes.

In this paper, we demonstrate {\it Castor}, a system for managing IoT time series data and predictive models along with semantic context information at scale.
The {\it Castor} architecture is composed of cloud-based microservices. {\it Castor} provides an efficient pipeline for real-time data ingestion, time series and model data management based on unified and intuitive application programming interfaces (APIs) to interact with the data and models. 
{\it Castor} employs Apache OpenWhisk \cite{Baldini2017} as its architecture framework and RabbitMQ \cite{videla2012rabbitmq} for a messaging fabric.
Deployment of models through OpenWhisk server-less cloud technology provides mechanisms for scheduling and scaling modelling workload (in particular, automated scoring and (re-)training), ultimately executed within containers \cite{merkel2014docker}.

From the Data Scientists' perspective, {\it Castor} supports the exploration of available time series data and predictive models, using high-level semantic concepts as entry points. The semantic context provides the Data Scientists with an immediate understanding of the structure of the domain and hence allows them to identify clusters of potentially related time series and/or hierarchical relationships that can be useful to guide model design and feature engineering.
Predictive models for deployment in {\it Castor} are created by implementing \textsc{load data},\textsc{transform data}, \textsc{train model} and \textsc{score model} functions. This can be done using the Data Scientist's preferred software (currently, Castor supports R and Python), and in their preferred 
computing environment (e.g., in a notebook on their local machine). Via the {\it Castor} APIs, the resulting predictive models (along with their semantic context information) can then be stored and deployed for automated scoring and (re-)training in the cloud production environment.

{\it Castor} continuously monitors all models in production and, based on user definitions, automatically schedules retraining jobs, for example upon performance deterioration, and scoring jobs, for example when new time series input data are available. Full traceability is ensured by maintaining a full history of the versions of trained models and of the time series predictions. Additional APIs can be used to explore existing models, their versions and prediction history or performance. Model reuse is also empowered, since the Data Scientist can easily identify existing models (possibly developed by other users) for signals or entities related to a new modelling task.

This paper is organized as follows. In Section II, we present the workflow of {\it Castor} from the Data Scientist's perspective. Section III discusses the architecture of {\it Castor} and individual microservices. Section IV demonstrates the functionality of {\it Castor} for a Smart Grid use case.

\section{{\it Castor} workflow} \label{sec:workflow}
In this section we describe the {\it Castor} workflow from a Data Scientist's perspective. As a running example throughout the rest of the paper we will consider a Smart Grid use case. The data is publicly available from the Open Power Systems Data Platform \cite{OpenData2017}. The time series that we aim to predict in this case are hourly measurements of electricity consumption and generation (in MWh/h) at various locations and aggregation levels in an electricity grid. External time series inputs used for the predictions include calendar features (e.g., time of day, time of year, season and day types), weather features (e.g., temperature, dew point, solar radiance) and autoregressive features (e.g., lagged energy consumption 24 and 48 hours ago). The semantic context information that is used to describe the Smart Grid domain is presented in more detail in Section IV. At this point we assume that the Data Scientist has identified the time series for which a predictive model shall be created, and all the relevant historical time series data have been ingested into the {\it Castor} system.

In order to create a predictive model that can be deployed for scoring and automated retraining in the {\it Castor} cloud production environment, the Data Scientist needs to implement a {\it Castor} model object, which consists of the following four functions: \textsc{load data},\textsc{transform data}, \textsc{train model} and \textsc{score model}. We are going to provide more details on those functions in the upcoming subsections. Note that, currently, {\it Castor} supports models implemented either in R or Python. For the demonstration in this paper, we created and deployed models in R.

Once the {\it Castor} model object has been created, it can be serialized and stored (along with its context information and deployment details) via the {\it Castor} APIs in the {\it Castor} model store. More details on the model store are provided in Section III.B, and more details on the deployment in Section III.D.

\subsection{Load data}
The entry point of the {\it Castor} model workflow is the \textsc{load data} step, where the user defines the required time series data to be loaded, both model covariates and target data. The time series data are identified using contextual information, as described more in detail in Section \ref{sec:data_management}. The {\it Castor} configuration parameters allow users to choose (1) the purpose of the retrieved data: training or scoring; (2) the length of the data. 

\subsection{Transform data}
Typical {\it Castor} \textsc{transform data} involves data transformation, anomaly detection and feature engineering. Basic transformation techniques are applied to stabilize variance or to enhance visualization, e.g., {\it log}, {\it square root} and {\it reciprocal}. Another important step in \textsc{transform data} is to detect and remove anomalies. Data anomalies are either due to faults in the telemetry and data warehouses or anomalous conditions in the operation. Hence, our approach includes: 

1) Outlier removal. Apply static rules to remove implausible values, such as negative values or constant segments.  

2) Change point detection. Iteratively apply the multiple change point detection algorithm, e.g., Pruned Exact Linear Time \cite{Killick2012}, to remove the segment which has the most significant deviation until the pre-defined threshold is achieved. 

3) Transfer learning (if applicable). Insufficient data might be left after handling change points. Transfer functions can be learned before and after the change points to re-align the data in order to enlarge the data size. 

The final step in \textsc{transform data} is to engineer additional features using the basic features  given in the data. It includes statistical features, e.g., daily min / max / average temperature, combination features, e.g., time of day with day type, solar radiance with season, and domain specific features, e.g, energy consumption peak hours and seasons.

\subsection{Train Model}
The {\it Castor} \textsc{train model} component can be flexibly defined by users, e.g., using decision tree or gradient boosting methods. Here we showcase a regression framework. 
\begin{eqnarray}\label{eq:gam_abstract}
Y_t  &=& \mu(X_t) + \sigma(X_t)\epsilon_t, 
\end{eqnarray}
where $Y_t$ is the hourly energy consumption measurement at time $t$ and $X_t$ is a vector of predictive features. The functions $\mu(\cdot)$ and $\sigma(\cdot)$ are the conditional mean and standard deviation, respectively, of $Y_t$ given $X_t$. The random variable $\epsilon_t$ is assumed to have zero mean and unit variance; it accounts for random fluctuations in $Y_t$. Given the predictive features $X_{t+h}$ for some future time point $t+h$, the conditional mean $\mu(X_{t+h})$ is our forecast of $Y_{t+h}$. Note that $\sigma(X_{t+h})$ can be used to quantify uncertainty in the forecast of $Y_{t+h}$. 

A number of options are available for the estimation of $\mu(\cdot)$ and $\sigma(\cdot)$. We follow the $\text{GAM}^2$ approach \cite{WSC2015, WAC+2016}, which applys two consecutively selected Generalized Additive Models (GAM)\cite{Wood2017} to respectively estimate $\mu(\cdot)$ and $\sigma(\cdot)$. GAM is a highly popular tool in the energy prediction applications due to its scalability, interpretability and high accuracy (see, e.g., \cite{Ba2012,GNK2014}).

1) ${\bf \text{GAM}^1 Step}$. The conditional mean $\mu(\cdot)$ is estimated using 
\begin{eqnarray}
\label{eq:gam}
\mu(X_t) &=& \beta + \sum_{j=1}^p f_j(X_t^{(j)})
\end{eqnarray} 
where $\beta$ is the model intercept, $X_t^{(j)}$ are 1- or 2-dimensional sub-vectors of $X_t$ (not necessarily disjoint), and the functions $f_j(\cdot)$ are cubic B-splines. E.g., the selected features in our example are $X_t^{\text{DayType}}$, $X_t^{\text{TimeOfDay}}$, $X_t^{\text{TimeOfYear}}$, $X_t^{\text{Temperature}}$, $X_t^{\text{SoloarRadiance}}$. Note that GAM models can handle both categorical and continuous features; in addition, features can be combined such that different spline functions are applied to continuous features depending on the values of categorical features. \\
\indent 2) ${\bf \text{GAM}^2 Step}$. Denote the estimated conditional mean by GAM as $\hat{\mu}(X_t)$. The squared empirical residuals $\hat{\sigma}^2(X_t)$ are obtained by 
\begin{eqnarray}
\label{eq:residual}
\hat{\sigma}^2(X_t) &=& \big(Y_t - \hat{\mu}(X_t)\big)^2. 
\end{eqnarray} 
We model $\hat{\sigma}^2(X_t)$ using another GAM, which includes features $X_t^{\text{TimeOfDay}}$, $X_t^{\text{DewPoint}}$, $X_t^{\text{DailyAverageTemperature}}$. 

The predictive features of GAMs are selected by the component-wise gradient boosting \cite{Buehlmann2007}, which is beyond the scope of this paper. For the feature selections and training of the GAM models, we use the {\tt mboost} and {\tt mgcv} packages in R (see \cite{Hofner2014, Wood2017}).

\subsection{Score Model}
After the set of models are trained, they will be stored into the {\it Castor} system. The model scoring and retraining will take place in the
cloud environment according to the user defined schedule or
upon the arrival of new data.

\section{System}
{\it Castor} is composed of a number of discrete, collaborating sub-components that operate natively on the cloud. We now highlight the architecture behind all components before discussing each of these components in turn.

\begin{figure}
  \centering
  \includegraphics[width=\linewidth]{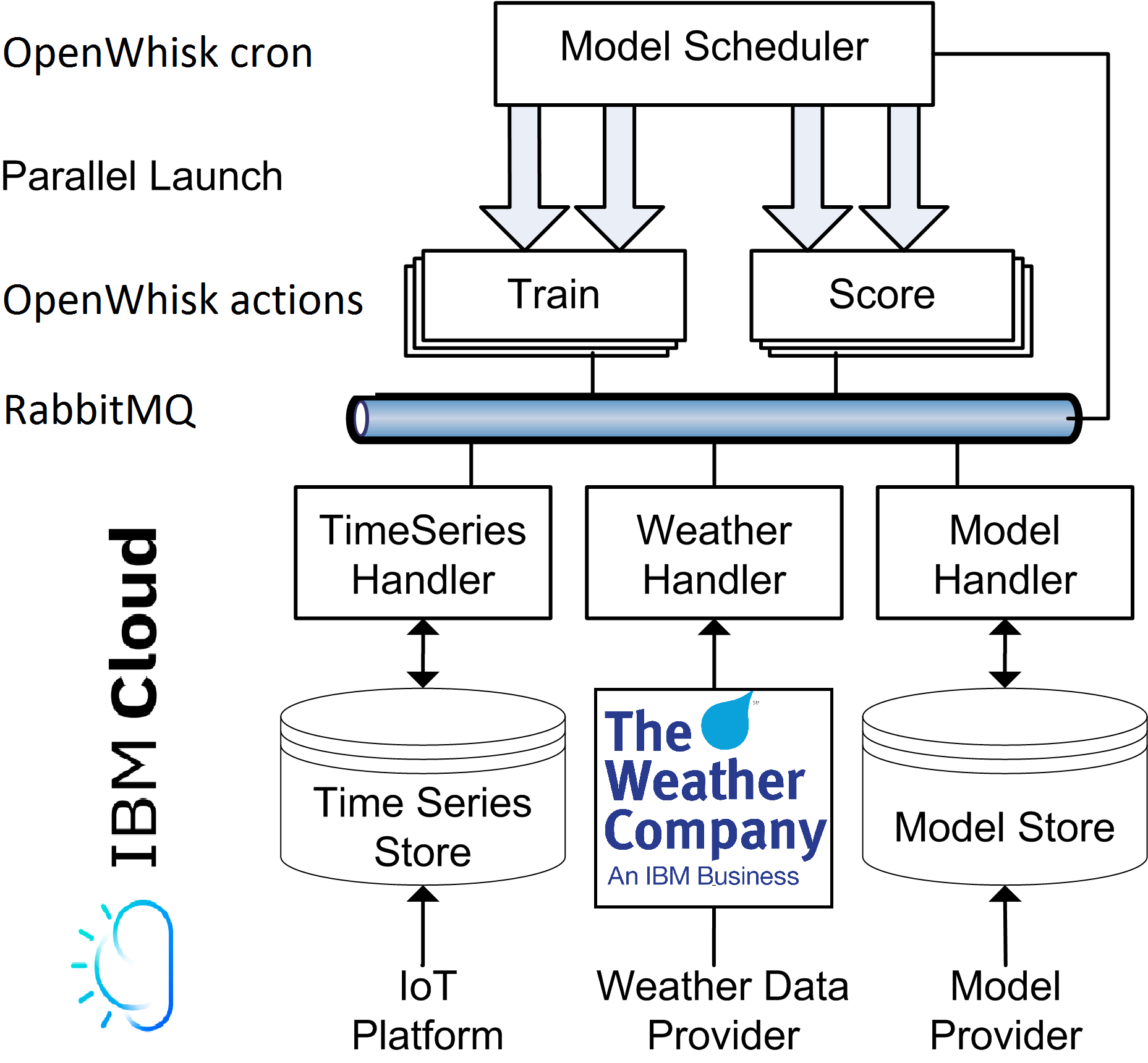}
  \caption{{\it Castor} architecture.}
  \label{fig:architecture}
\end{figure}

\subsection{Architecture}
{\it Castor} follows a microservices architectural pattern. Each individual microservice is well defined and performs a specific task, such as storing newly acquired time series. The microservices pattern is quite different to traditional architectures, which often produce monolithic applications. In contrast, microservices are loosely coupled, and are accessed via lightweight, well-defined APIs. These APIs are based on asynchronous messaging protocols.

Fig. \ref{fig:architecture} shows the various microservices and their coupling in {\it Castor}. Time series data coming from the models (i.e.~predictions and forecasts) or directly from the sensors are stored in the {\it Time Series Store}. In addition, the system is capable of directly ingesting data from external providers such as weather data. Models created by Data Scientists are stored in the model store. These two stores are linked through their common context for the Data Scientist to easily access and manipulate models or data as required. After the models are stored, the model scheduler handles the automation of the training and scoring of these models using OpenWhisk actions and Advanced Message Queuing Protocol (AMQP) messaging based on the default or user defined deployment configurations.

\subsection{Time series and model data management with Context}
\label{sec:data_management} 

The elementary objects persisted in the {\it Castor} Data Store are:
\begin{enumerate}
\item \emph{Time Series}, represented as raw time sequences of numeric or character values.
\item \emph{Models}, represented as binary representation of the custom user code performing the load / transform / train / score workflow discussed in Section \ref{sec:workflow}. Additional metadata contain a name, a description and the desired training schedule. Models can have multiple \emph{model versions}, which store the results of each training job (e.g. model parameters) and have their own scoring schedule. Training and scoring schedules allow the user to control automatic deployment of training and scoring jobs on the cloud through the model scheduler component, detailed in Section \ref{sec:model_scheduler}.
\end{enumerate}

\begin{figure}[H]
  \centering
  \subfigure[] {
  \includegraphics[width=0.4\linewidth]{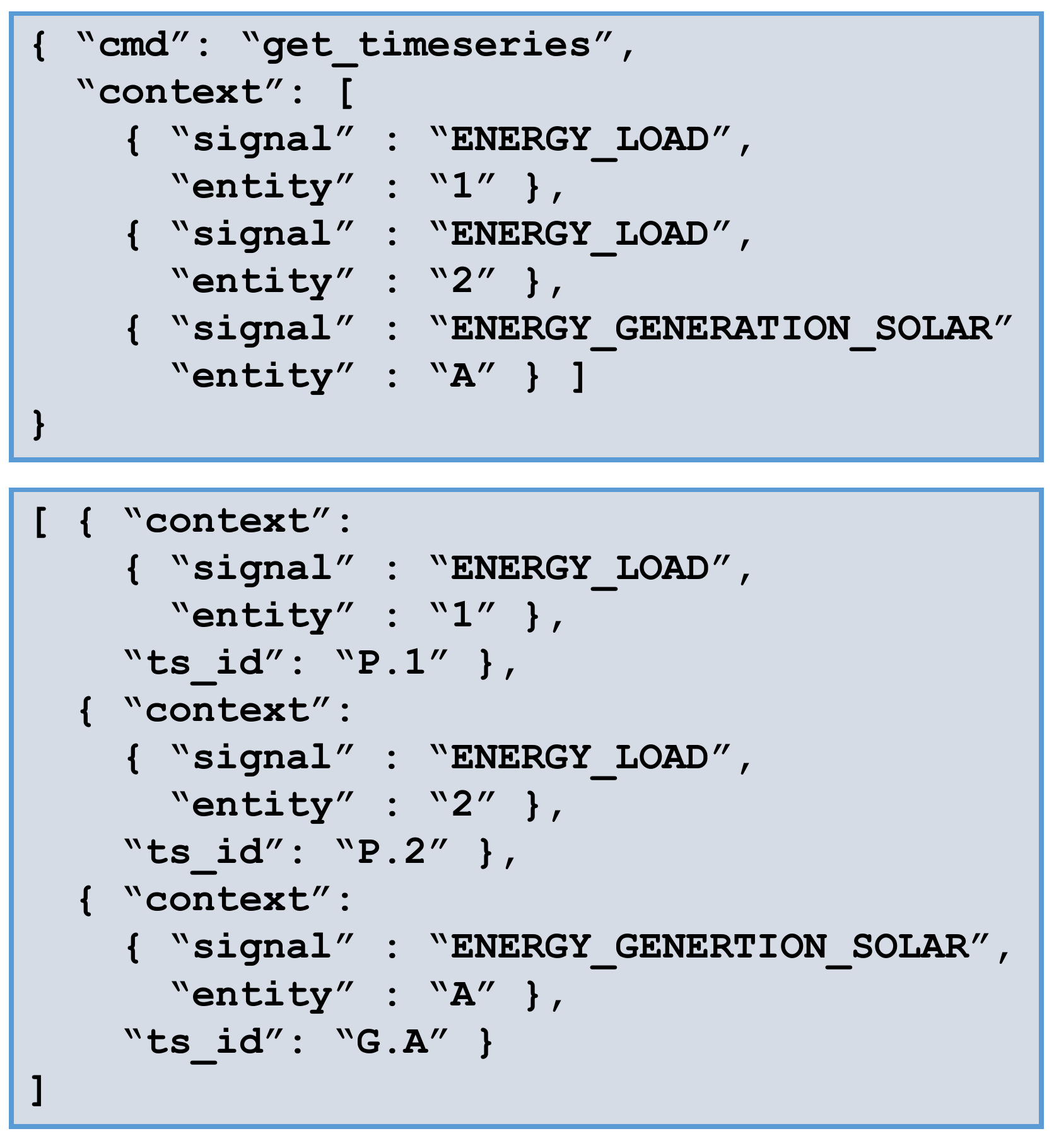}}
  \subfigure[] {
  \includegraphics[width=0.4\linewidth]{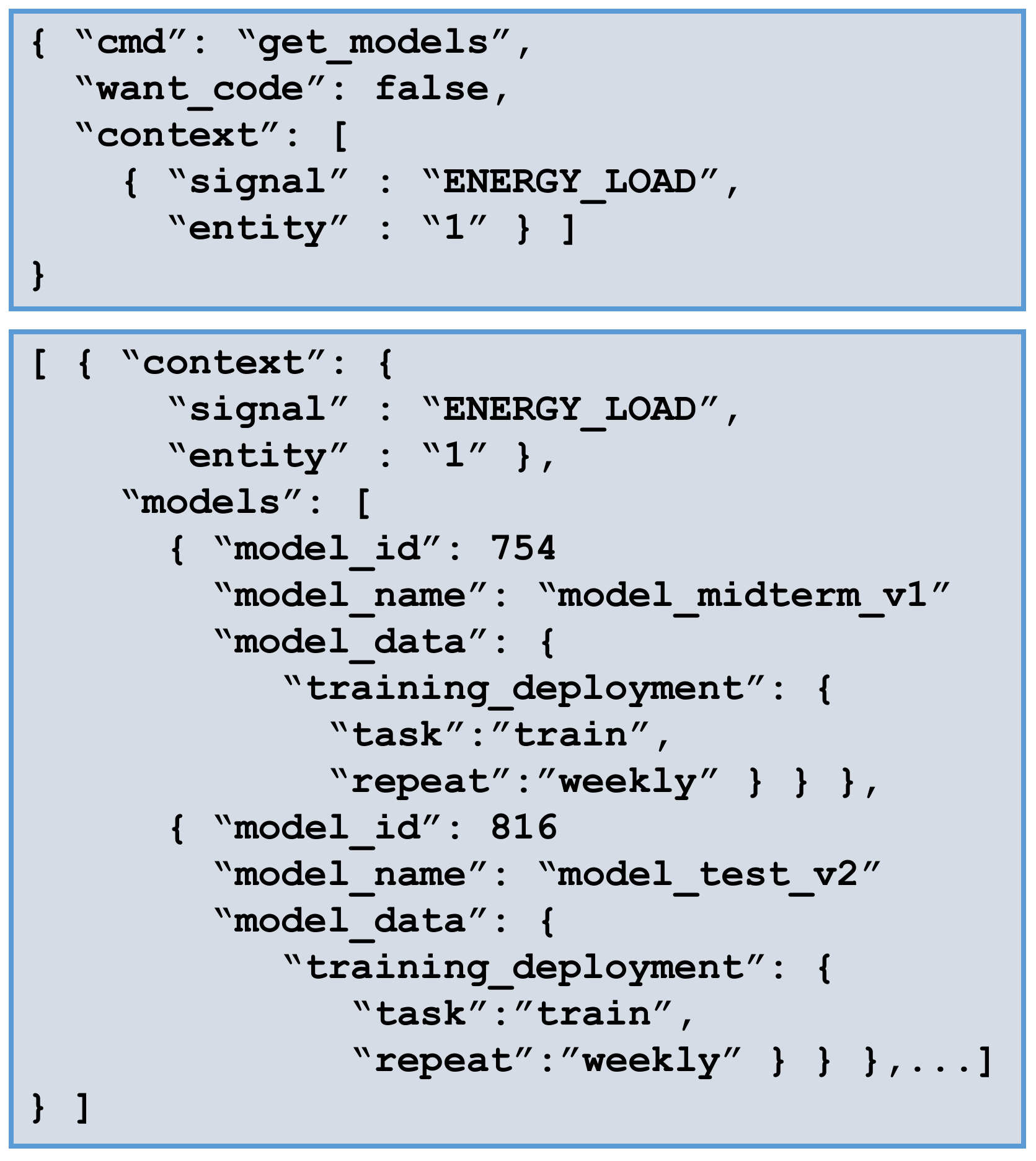}}
  \caption{{\it Castor} API sample requests (top) and responses (bottom) for: \emph{(a)} time series; \emph{(b)} models.}
  \label{fig:castorapi_sample}
\end{figure}

\begin{figure}
  \centering
  \includegraphics[width=\linewidth]{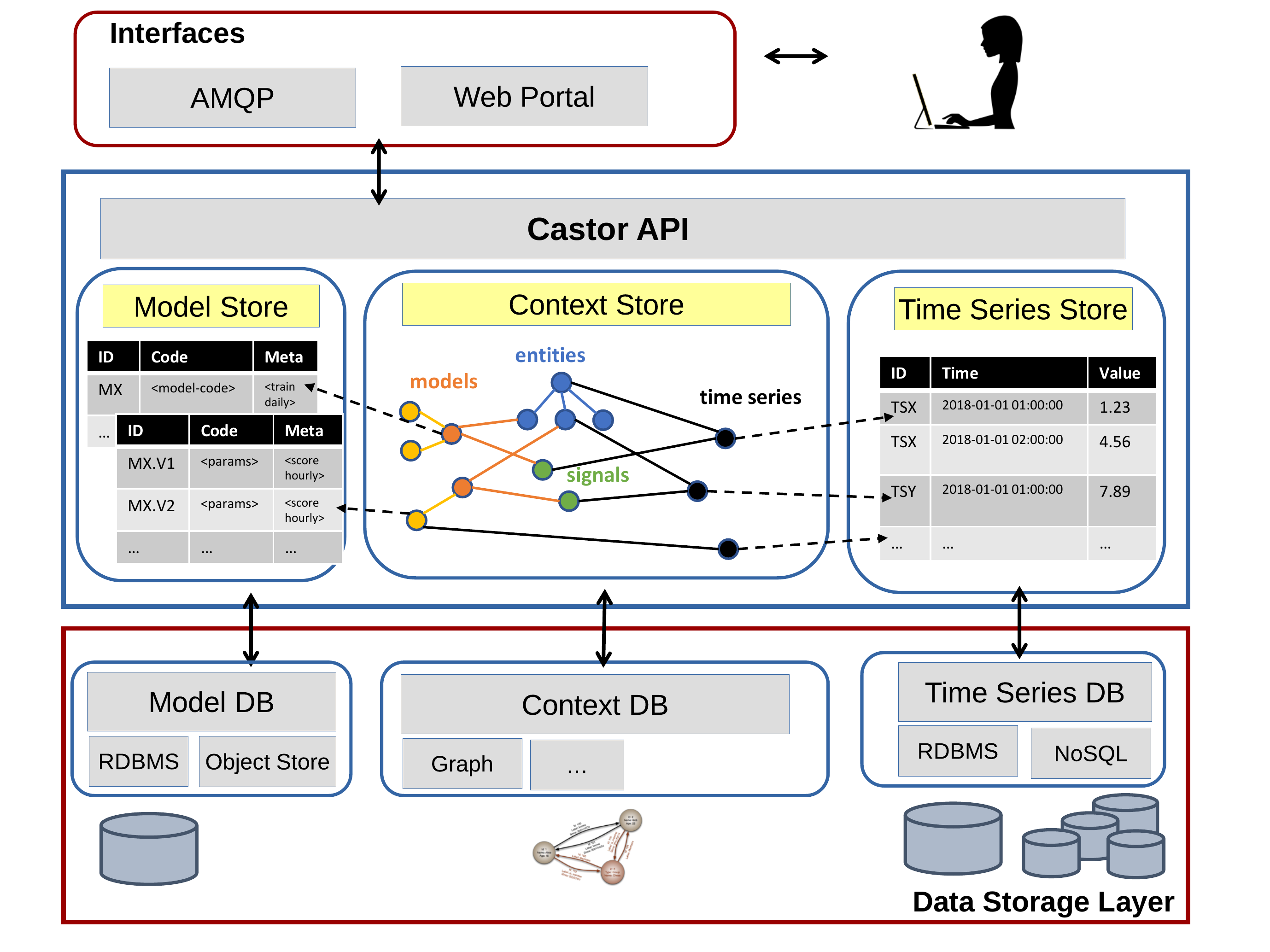}
  \caption{{\it Castor} time series and model data management component diagram.}
  \label{fig:data_store}
\end{figure}

Besides pure time sequences, sensor networks often come with meta data providing a rich semantic description of the application domain where the time series reside in. Such meta data define a \emph{context} representation of the relationships of the time series with domain entities or with other time series, which can be particularly complex for IoT applications due to the potential heterogeneity of the metering data sources. It is important that intelligent data-driven applications have seamless access to the time series data along with their context information, such to exploit their full potential for gaining insights or taking decisions \cite{Ploennigs2017,Goynugur2017,Schumilin2017}.

The approach to data management in {\it Castor} is therefore based on the idea that both time series data and models are handled consistently with their context data. Context is defined as a combination of an \emph{entity} and a \emph{signal}. A signal defines any physical quantity measured over time (e.g. temperature, hourly electrical energy consumption, daily ice cream sales). An entity is any location of the domain that can be associated with signals (e.g. a town, a customer, a shop) and can have a geography. Both signals and entities are conceptually grouped by the concepts of signal and entity types, respectively. Hierarchical and topological relationships between entities can also be defined, thus allowing for a very general framework for defining complex semantic domain descriptions. As shown in the demonstration (see Section \ref{sec:demo}, context data is key in supporting the Data Scientist in data exploration, curation, feature and model engineering tasks. Furthermore, by leveraging the high-level context-based {\it Castor} APIs (see samples in Fig. \ref{fig:castorapi_sample}) automation of data curation processes, feature and model selection, or even automated creation and replication of predictive models are possible.

In order to deal both with the highly relational nature of context data and the (potentailly large-scale) tabular data required for timeseries and analytical models, the system makes use of a multi-modal data store in the back-end. Time series and model data are stored in a traditional RDBMS. Context data, including relationships with models, model versions and time series, are managed in a graph database, which is particularly well-suited for handling data elements that have rich and complicated relationships \cite{Zhang2017}. A high-level {\it Castor} API encapsulates the complexities involved in serving high-level intuitive user queries for storing or retrieving time series data or model data based on context by navigating the different stores as needed. The {\it Castor} data store is designed to be transparent to the specific database technology used in any of the time series, model or context stores used in the back-end. No-SQL database technologies (Cassandra, HBASE) can be alternatively used for storing time series data, or object stores could be used to store the models. A component diagram of the data management architecture behind {\it Castor} API is shown in Fig. \ref{fig:data_store}.

\subsection{Modelling infrastructure}
\label{sec:model_infra}
To support the training and scoring of a varied set of models, we employ the serverless \cite{Baldini2017} Apache Openwhisk framework. Running on the Cloud, each individual function is referred to as an action. The server-less attributes remove the burden of server-side resource management and provide built-in mechanisms for both scheduling and scaling action invocations. Using the action scheduler (similar to cron), we trigger our model scheduler to regularly examine model training/scoring schedules, and launch appropriate downstream actions when required. Typically, many model training/scoring actions are launched in parallel; with the launcher exiting when all model schedules are satisfied. Model training/scoring actions trigger the execution of a docker container that retrieves the model code (code and core) and stores the model version (forecasts) on successful completion of the task.

While OpenWhisk actions orchestrate the execution of training and scoring jobs, messaging plays an important role in storing and retrieving models and time series data. We use RabbitMQ \cite{videla2012rabbitmq} as our messaging fabric, from which each launched model action requests training/scoring data (time series, weather observations and forecasts). Upon completion of a training/scoring action, a notification is published to RabbitMQ, enabling subsequent services to be triggered.

\subsection{Model scheduler} \label{sec:model_scheduler}
\begin{figure}
  \centering
  \includegraphics[width=\linewidth]{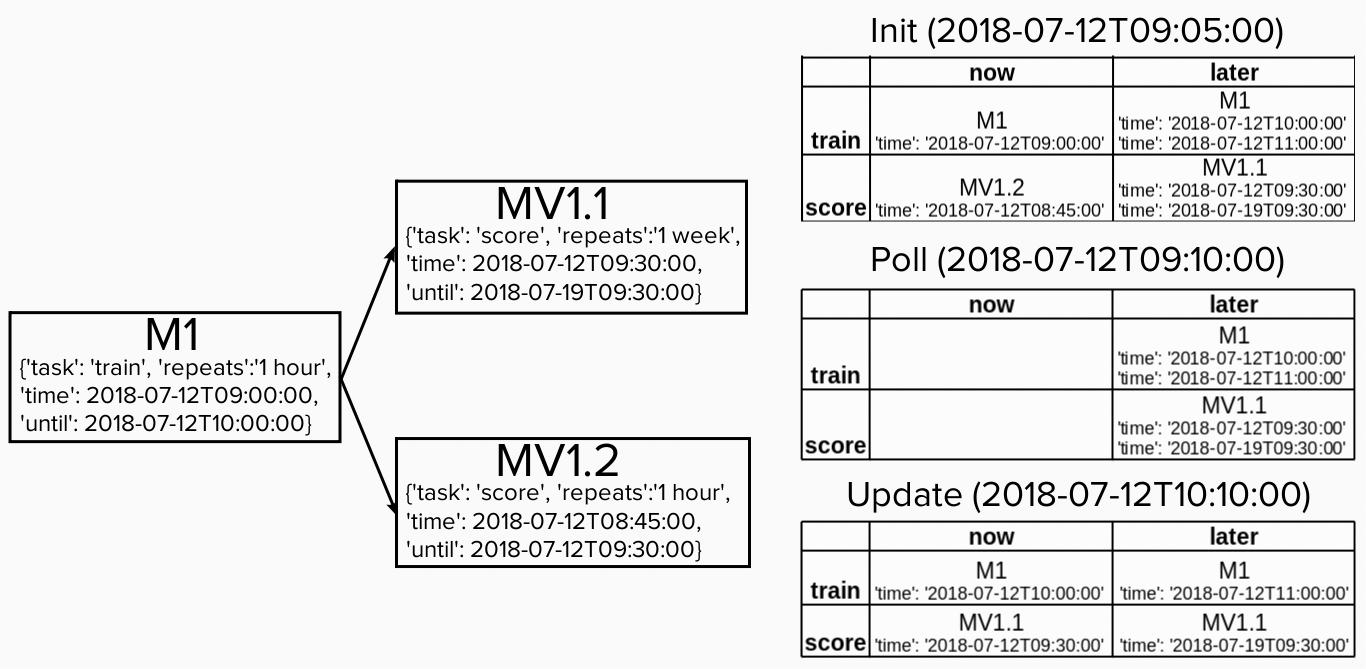}
  \caption{Illustration of the interaction between model store, openwhisk actions and redis queues through model scheduler.}
  \label{fig:model_scheduler}
\end{figure}

The Model scheduler is a service to trigger the training and scoring of models and model versions (trained models) described in Section \ref{sec:data_management}.  The Data Scientist provides the training and scoring deployment configuration for every model, and this information is stored in the database with the model. An example of the training configuration is shown in Fig. \ref{fig:model_scheduler}. The training cofiguration provides information on the task that needs to be performed (`train' or `score'), the time to perform the task for the first time, the frequency of the task (`repeat') and the end time of the task (`until').

The Model scheduler works on three different cloud functions:

{\bf Init} action which uses the deployment configuration set by the user and brackets models and model versions from the database into Redis task queues \cite{carlson2013redis}. There are two queues each for training and scoring, namely `now' and `later'. 
If the `time' in the  `deployment configuration' is past the `time' when the cloud function is performed, the model is set in the `now' queue and `later' otherwise.

{\bf Poll} action which scans the redis ``now'' queues and launches the docker container to run the model code for training and scoring as defined in the task queue. The workflow of the docker container is mentioned later.

{\bf Update} action which updates the task queues on a regular basis by moving tasks from the ``later'' to the ``now'' queues when they are ready to be deployed based on the invocation time as described earlier for the ``Init'' action.

We illustrate these actions with the help of an example. In Fig. \ref{fig:model_scheduler}, we consider one model M1 and two model versions MV1.1 and MV1.2 with the corresponding deployment configurations to be present in the model store. The ``Init'' action is performed at 9:05 AM on 2018-07-12. Since M1 and MV1.2 have deployment configurations before the ``Init'' action time, they are placed in the `now' queues. The remaining possible jobs are determined by taking the `repeat' field in the deployment configuration and placing the jobs in the `later' queues. The ``Poll'' action is invoked at 9:10 AM on 2018-07-12. This scans the redis `now' queues and launches the docker containers for M1 and MV1.2 respectively. This action results in the creation of a new trained model (from M1) and a new forecast (from MV1.2) which are stored in the model store and timeseries database respectively. The ``Update'' action is invoked at 10:10 AM on 2018-07-12 and it moves one task from M1 and MV1.1 to the `now' queues since two tasks in the redis `later' queues are before the ``Update'' invoke time. The cron invocations of these three Openwhisk actions thus ensure that the model and and model versions are automatically trained and deployed at user-defined intervals.
The Model scheduler invokes the model docker container through the ``Poll'' cloud function to deploy the model. The workflow for the container is as follows:

1) The Model container receives a model id or model version id, along with the information whether this is a training or scoring job from the ``poll'' cloud function. It also receives the timestamp at which this job was due (which must have elapsed, otherwise the container would not have been triggered to execute the job).

2) The Model container uses the model id or model instance version id to query from the Model Store (via AMQP) all the information that is required to execute the job. In particular, if the task is ‘train’, then the container will retrieve the model code and if the task is ‘score’, then it will retrieve the model code and model instance version core.

3) The model container then executes the following sequence: \newline
{\bf Load data}: The timestamp at which this job was due can be used for determining the exact window of data to be loaded (the user would have been responsible for implementing this logic in the model code). Data can be loaded from the TimeSeries Store and/or the Weather Data Service via AMQP. \newline
{\bf Transform data}: Apply transformations of the data implemented by the user in the model code. \newline
{\bf Train model}: (if the task is ‘train’), serialize it and store it as a new model instance version in the Model Store (via AMQP). If model training was not successful, create an error log message.\newline
{\bf Score model}: (if the task is ‘score’), and store the output in the TimeSeries Store (via AMQP). If model scoring was not successful, create an error log message.

\subsection{Scalability}

A technical challenge of time series modelling in production environments is frequent re-training and scoring of predictive models on receipt of new data. For example, to maintain accurate forecasts of renewable energy production and consumption, model predictions should be regularly updated using the latest observed weather and recorded production/consumption data, typically every few minutes. Similarly, the pace of change in the environment, e.g. new solar generation installations or changes in customer electrical consumption patterns, require regular benchmarking and retraining of predictive models.

In order to cope with such time-critical, \emph{bursty} model training and scoring compute workloads, {\it Castor} adopts dynamic scaling  \cite{Jonas2017}, whereby additional compute resource is scaled up to handle bursty workloads, and scaled back once the compute demand peak has passed.
As described in Section \ref{sec:model_infra}, this is accomplished through the OpenWhisk cloud service.
Its scalability characteristics allow up to a thousand actions, or logical compute nodes, to be provisioned in a few seconds, each capable of running a time-series modelling training or scoring job. 
 Its elastic design minimises technical complexity as no additional infrastructure code is needed to configure and manage scaling of underlying compute resource.
Table \ref{tab:scalability} shows the horizontal scalability achieved (and projected) by the current Castor infrastructure with respect to scoring and training jobs on the cloud. The number of concurrent jobs depends on the amount of resources (and cost) allocated for the OpenWhisk service.

\begin{table}
\centering
\begin{tabular}{|c||c|c|}
\hline
 \# Concurent Jobs & Ave Duration [s]  & Max \# Jobs per hour  \\
\hline 
\hline
 25 & 180 (60)  & 500 (1500) \\
 50 & 180 (60)  & 1000 (3000) \\
 100$^*$ & 180 (60)  & 2000 (6000) \\
 500$^*$ & 180 (60)  & 10000 (30000) \\ 
\hline
\end{tabular}

\caption{Scalability of predictive model training (scoring) jobs with Castor. ({\it $^*$ projected })}
\label{tab:scalability}
\end{table}

\section{Demo}
\label{sec:demo}
\begin{figure}
  \centering
  \includegraphics[width=\linewidth]{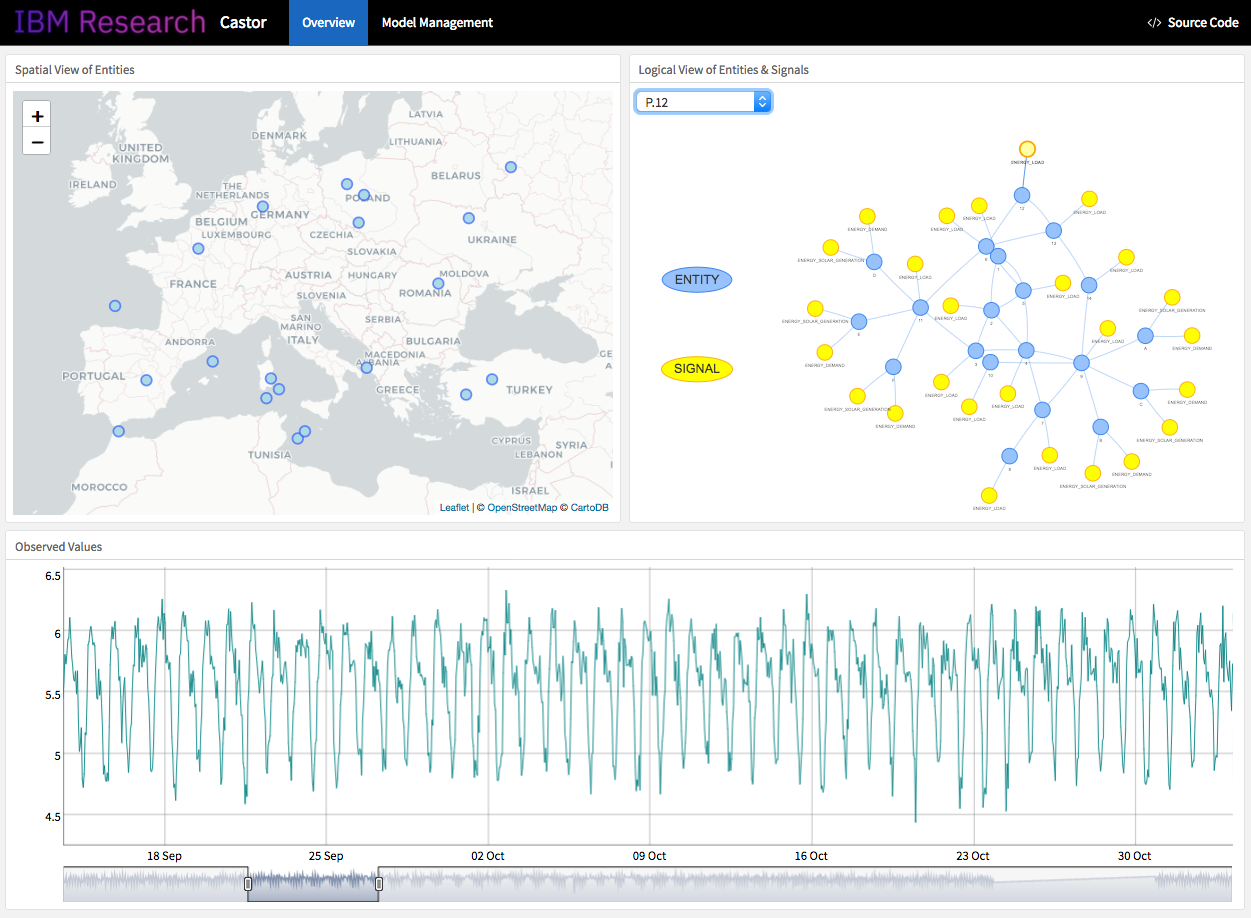}
  \caption{Overview screen showing spatial location of entities and context view with layers of entities, signals and time series.} 
  \label{fig:demo-overview-tseries}
\end{figure}

In this section, we present a key features and functionalities of the {\it Castor} Demo. Our demonstration is a web application built using the Shiny \cite{Chang2018} framework for the R environment. It utilizes a custom R package that implements {\it Castor}'s JSON-based AMQP APIs. As shown in Fig. \ref{fig:demo-overview-tseries},  the demonstration includes {\bf Overview}, {\bf Time Series} view and {\bf Model Management} view. 

{\bf Overview}.  As discussed in Section \ref{sec:data_management}, {\it Castor} relates both time series data and models to their context, which is represented by a pair of entity and signal. The {\bf Overview} of our demonstration, shown in Fig. \ref{fig:demo-overview-tseries}, provides a side-by-side view of the spatial and topological distributions of all the entities in the system. On the left-hand side, the existing entities are clearly marked on a map while on the right hand side, the entities and signals are represented in a graphical model. Users can further explore the context by selecting the time series layer, which expands the graphical model in Fig. \ref{fig:demo-overview-tseries} with the relationships between the time series data and the corresponding entities and signal. The purpose of {\bf Overview} is to provide users with a general understanding of the information stored in the {\it Castor} system and how they are organized. The hierarchical structure and dependency cluster visualized in this layer can be extremely valuable to users for feature engineering and predictive model design.

The bottom of the overview page shows a visualization of the time series selected from the context graph. The entire time series available in the {\it Castor} system will be shown as an interactive plot. To focus on a segment of the chosen time series, the range of data can be narrowed / adjusted using a slider control. Hovering over the graph shows values for individual points.

\begin{figure}
  \centering
  \includegraphics[width=\linewidth]{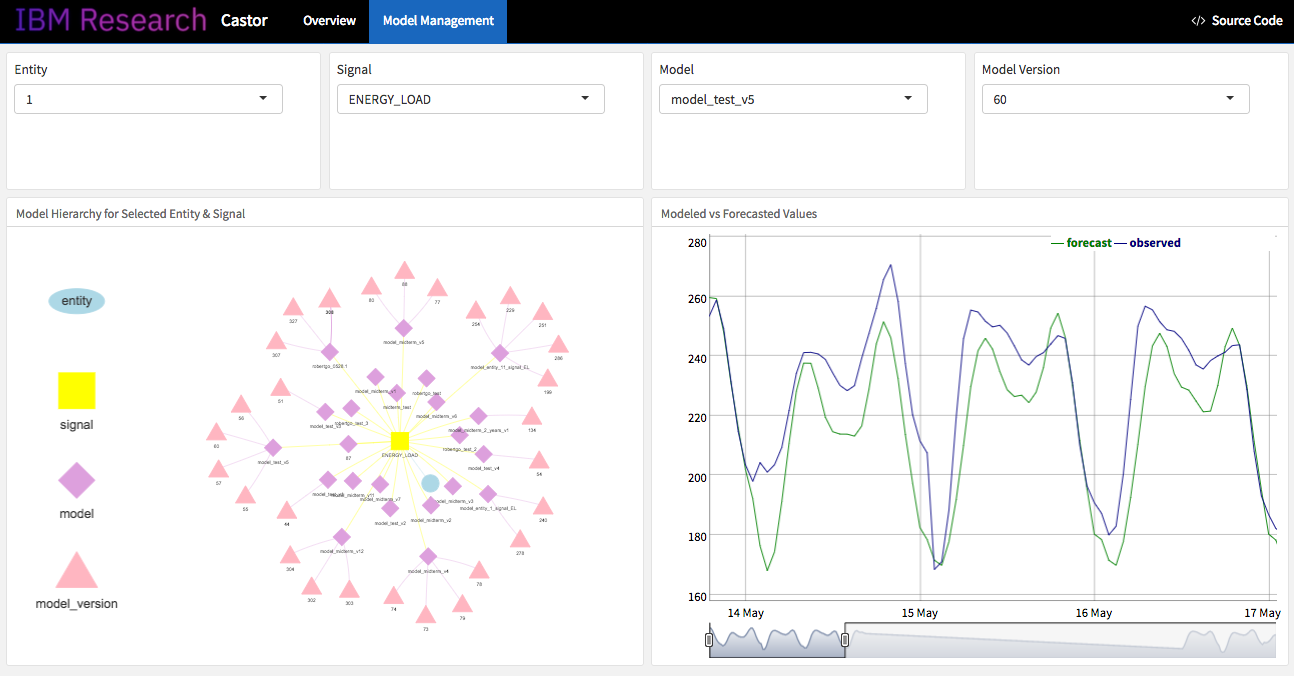}
  \caption{Model management screen showing hierarchy of models and model versions along with forecast and observed values.}  
  \label{fig:demo-modelmanagement}
\end{figure}

{\bf Model Management}. The {\bf Model Management} view provides interactive exploration of the models available for the selected entity and signal, shown in Fig \ref{fig:demo-modelmanagement}. The graphical model of the context can be expanded with the model layer. This option provides additional information on which signal and entity the model associated with. When creating models for a new context, users can utilize this view to easily locate relevant models (possibly created by some other users) which could be reused because they target at the same type of signal at an entity close by. Topological and hierarchical relationships can also be leveraged by users to inform transfer learning or model boosting tasks where several models are used to create a model for a related entity.

One of the key features of our system is the ability to maintain multiple models and model versions. Here the model refers to a representation of the time series data, such as artificial neural network, GAM, random forest and XGBoost. The model version refers to an instance of such a model that has been trained on a particular dataset. The default behaviour in the system is to run the latest version of the model.  On the left hand side of {\bf Model Management} is a network view of the model hierarchy showing the models and model versions that are available for a selected entity and signal; On the right hand side is a comparison between the selected model and model version with observed values for the chosen signal. The {\bf Model Management} view helps users to check the current model accuracy, track different versions of models and easily alternate among the models with the best predictive performance. 

\section*{Acknowledgements} 
This research has received funding from the European Research Council under the European Union’s Horizon 2020 research and innovation programme (grant agreement no. 731232). Contributions to an early version by Francesco Vigliaturo are gratefully acknowledged.

\balance
\bibliographystyle{abbrv}
\bibliography{goFlexBib}  

\end{document}